\documentclass[12pt,oneside,draftcls,onecolumn]{IEEEtran}%
\usepackage{amssymb}
\usepackage{amsmath}
\usepackage{amsfonts}
\usepackage[dvips]{graphicx}
\usepackage{color}
\usepackage{cite}%
\usepackage{amscd}
\usepackage{balance}%
\usepackage{graphicx}

\newtheorem{theorem}{Theorem}

\begin{document}
\bstctlcite{IEEEexample:BSTcontrol}
\markboth{IEEE Transactions on Communications, VOL. XX, NO. X, XXXX
XXXX}{Bithas, \textit{et al}: UAV-to-Ground Communications: Channel Modeling and UAV Selection}

\title{UAV-to-Ground Communications: Channel Modeling and UAV Selection}
\author{Petros~S.~Bithas,~\IEEEmembership{Senior Member,~IEEE}, Viktor~Nikolaidis,~\IEEEmembership{Student Member,~IEEE}, Athanasios~G.~Kanatas,~\IEEEmembership{Senior Member,~IEEE}, and George~K.~Karagiannidis,~\IEEEmembership{Fellow,~IEEE}  
\thanks{Manuscript received November 7, 2019; revised February 13, 2020; accepted April 12, 2020. Date of publication XXXXXX XX, 2020; date of current version XXXXX XX, 2019. This research has been partially funded by project BEAM of the University of Piraeus Research Center.}
\thanks{P.~S.~Bithas is with the General Department, National and Kapodistrian University of Athens (e-mail: pbithas@uoa.gr).}
\thanks{V. Nikolaidis and A.~G.~Kanatas are with the Department of Digital Systems, University of Piraeus, 18534 Piraeus, Greece (e-mail: {vnikola;kanatas}@unipi.gr).}
\thanks{G. K. Karagiannidis is with the Aristotle University of Thessaloniki, Thessaloniki 54636, Greece (e-mail: geokarag@auth.gr).}
\thanks{Digital Object Identifier 10.1109/XXXXXXXXXXXXXXX}
}

\maketitle
\vspace{-2cm}
\begin{abstract}
Unmanned aerial vehicle (UAV)-enabled communications have been proposed as a critical part of the beyond fifth-generation (5G) cellular networks. This type of communications is frequently characterized by line-of-sight (LoS) and dynamic propagation conditions. However, in various scenarios, the presence of large obstacles in the LoS path is unavoidable, resulting in shadowed fading environments. In this paper, a new channel model is proposed, in which the effects of mobility and shadowing are simultaneously considered. In particular, the performance of a UAV-based communication system operating in a shadowed double-scattering channel is analyzed. {The new channel model is generic, since it models various fading/shadowing conditions, while it is in terms of easy-to-evaluate mathematical functions. Moreover, a low complexity UAV selection policy is proposed, which exploits shadowing-related information. The proposed scheme offers a reduction of the signal processing complexity, without any important degradation on the performance, as compared to alternatives approaches.} In this context, a new analytical framework has been developed for investigating the performance of the new strategy. Finally, the main outcomes of this paper are also validated by empirical data, collected in an air-to-ground measurement campaign. 
\end{abstract}%
\vspace{-0.5cm}
\begin{IEEEkeywords}
Air-to-ground measurement campaign, composite fading, double scattering, goodness-of-fit tests, unmanned aerial vehicles (UAVs) selection (UAV association).
\end{IEEEkeywords}

\section{Introduction}
Unmanned aerial vehicle (UAV)-enabled networks have been proposed as a critical component of the beyond fifth-generation (5G) networks, to satisfy the requirements for massive connectivity, ultra reliability, and increased throughput. By operating as aerial user equipments (UEs) or as flying base stations (BSs), UAVs provide rapid recovery of the network services as well as system offloading in {crowded} situations. This technology has already attracted the interest of academia \cite{7470933,7463007,8657707}, industry \cite{russon2016nokia,2017huawei}, and standardization bodies \cite{3GPP_UAVs}. However, although UAVs add a new degree of freedom to the network, they have also induced various new challenges regarding physical layer communication aspects, such as channel modeling, UAV selection strategy, placement/trajectory planning, and interference management. This paper attempts to investigate the first two research challenges.

In UAV-to-ground communications, the wireless medium is characterized by high mobility and line-of-sight (LoS) propagation conditions \cite{7108163,8579209}. However, in several cases, due to the presence of large obstacles, the LoS assumption is not satisfied \cite{8411465}. In such scenarios, {shadowing (or \textit{large-scale} fading) occurs, which results in randomly vary mean envelope levels}. Another independent phenomenon, characterizing the mobile communication environments, is the \textit{double-scattering (DSc)} propagation. This type of fading is observed in scenarios where the transmitter (Tx), the receiver (Rx), or important scatterers around them, are in motion \cite{andersen2002power,salo2006statistical}. In this environment, {the transmitted signal propagates via multiple reflections, scattering, and diffractions around both the local scattering regions of the Tx and the Rx. Thus, two independent sums of plane waves are multiplied to form the resulting impulse response, originating what is known as DSc model \cite{andersen2002power,salo2006statistical,8052560}. However, the combination of DSc and shadowing has rarely been investigated in the past. For example, in \cite{5426028}, Nakagami-$m$ distribution was assumed for the small-scale fading and gamma for the shadowing. From the psychical point of view, this shadowed DSc channel is a reasonable approach to model the random variations of the envelopes' mean values.}

Recently, the inverse-gamma (IG) distribution has been proposed as a shadowing model, since it has been found to offer an excellent fit to measurement data for large-scale fading, e.g., \cite{8166770,8377390,Bithas_PIMRC_2018,WCNC2019}. In particular, in \cite{Bithas_PIMRC_2018}, it was shown that the IG distribution offers a better fit (as compared to the gamma) to empirical data for shadowed fading conditions in non-stationary conditions. In \cite{7343288}, it was also shown that the $\eta-\mu$/IG distribution provides a good fit to shadowing modeling in body-centric communications. Finally, in \cite{WCNC2019}, a new shadowed DSc model for UAV-to-ground communication was presented, which was based on the Nakagami-$m$ and IG distributions for modeling the multipath fading and shadowing, respectively. In that paper, it was shown that this new distribution provides an excellent fit to empirical data obtained in air-to-ground (A2G) measurement campaign.

The performance of UAV-enabled communications has been analyzed in various studies in the past, e.g., \cite{5937283,8647521,8501925,8017572,8594719,8733808,8765633,8401901,8713514}. Specifically, in \cite{5937283}, a communication system was investigated, in which UAVs were used as relays between ground terminals and a BS. In the same paper, a performance optimization algorithm was developed, through the control of the UAV heading angle. In \cite{8647521}, the outage probability (OP) of UAV-based communications has been analyzed by taking into consideration the {co-channel interference, from aerial or ground interfering nodes,} in the case of LoS and non-LoS (NLoS) links. In \cite{8501925}, the OP is analyzed in a UAV-assisted relaying system operating in a composite fading environment, modeled by the shadowed-Rice distribution. In \cite{8594719}, a multi-hop UAV relaying system has been analyzed, by taking into account the influence of both multipath and shadowing effects. For this system, the adopted performance metrics were the bit error rate and the channel capacity. In \cite{8733808}, considering a Rician-shadowed environment, the OP of hybrid-duplex UAV communication system was analyzed. In \cite{8401901}, the performance of a low-altitude aerial relaying platform is also investigated in terms of the OP. 

{As far as UAV-selection policies are concerned, various approaches have been proposed in the past, e.g., \cite{7842359,8972910,8713514,8917591}. More specifically, in \cite{8713514}, among other research targets, a UAV selection policy was proposed, in which the UAV offering the highest instantaneous received power was selected. In \cite{7842359}, a UAV-selection mechanism was proposed, based on two linear integer optimization problems, targeting to minimize the energy consumption and the operation time. In \cite{8972910}, three UAV relay selection strategies were proposed, namely  the closest-UAV relay selection, the maximum signal-to-noise ratio (SNR) UAV relay selection, and a random selection, with the second one offering the best performance with the cost of high channel state information (CSI) feedback and complexity. The maximum SNR rule was also adopted in \cite{8917591} as a relay selection policy. Nevertheless, such a policy offers a good performance with the cost of increased overhead and signal processing complexity, since a continuous monitoring of all available links is required.}

Motivated by the above, in this paper, a new shadowed DSc distribution is proposed, which is able to model situations where the received signal is simultaneously affected by DSc and shadowing. In this context, the performance of a communication system operating in the aforementioned environment is analyzed, employing well-known metrics, i.e., bit error probability (BEP), channel capacity, and OP. Additionally, a new UAV-selection policy is proposed that exploits the information related to the shadowing. For the proposed scheme, the probability density function (PDF) and the cumulative distribution function (CDF) of the output SNR are provided in closed form, while its performance is evaluated using the criteria of OP and average output SNR (ASNR). Finally, simulated results and empirical data\footnote{The data were collected in an A2G measurement campaign, which involved a zeppelin-type airship \cite{nikolaidis2017dual}.} are also exploited to validate the analytical results and prove the applicability of the proposed approach in real-world communication conditions. Consequently, the main contributions of this paper are threefold:
\begin{itemize}
    \item A new composite fading distribution is proposed that simultaneously models the impact of DSc and shadowing on a UAV-enabled communication system. {The new distribution asymptotically includes as special cases previously reported ones, while it is given in a mathematical convenient form};
    \item A new UAV-selection policy is proposed, {which exploits the stationarity of the shadowing. The new scheme offers similar performance with alternative approaches, which are based on the instantaneous CSI, with an important reduction on the signal processing overhead};
    \item Empirical data, {collected in UAV-to-ground measurements campaign,} are employed to support the proposed {i) channel model and ii) UAV selection policy.}
\end{itemize}

The remainder of the paper is organized as follows. In Section II, the shadowed double-scattered channel model is presented. In Section III, the performance of a single communication system operating over the considered channel model is analyzed. In Section IV, the new UAV-selection strategy is presented and its performance is analyzed. In Section V, empirical results and comparisons are presented and discussed, while in Section VI, the concluding remarks can be found.

\begin{figure}
\centering
\includegraphics[keepaspectratio,width=4in]{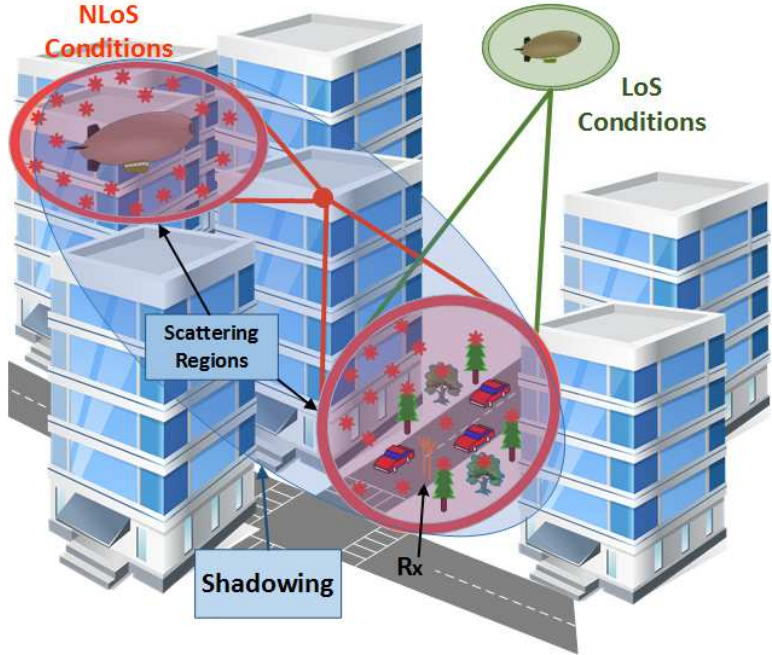}
\caption{The communication scenario under investigation.}\label{fig:A4_Scenario}
\end{figure}

\section{System and Channel Model}
We consider a communication scenario in which the Tx is located in a UAV and the Rx is located at the ground. It is also assumed that the Tx and/or the Rx and/or important scatterers around them are in motion, while the two local regions of scatterers, around both Tx and Rx, are separated by a large distance. Such a communication environment is expected to result in a double-scattered received envelope, due to the keyhole propagation \cite{salo2006statistical}. Moreover, it is reasonable to assume that shadowing exists due to the presence of obstacles between the Tx and the Rx, e.g., buildings. Therefore, in the considered channel model, large scale fading (or shadowing) co-exists with the double-scattered propagation. A representative example of the communication scenario under investigation is shown in Fig.~\ref{fig:A4_Scenario}. {In this figure, two UAV-enabled communication scenarios have been illustrated, namely one with LoS conditions and one with NLoS, while the two scattering regions as well as the shadowing area have also been identified}.

\subsection{Double-Shadowing Communication Scenario}
Next, we investigate the double-shadowing (DS) scenario, which is the case when shadowing {(and as a consequence NLoS conditions) is present in both local scattering regions at the Tx and the Rx}. Let $\gamma_i$ denoting the received SNR from the $i$th (with $i\in\{1,\dots, L\}$) UAV, which is defined as
\begin{equation}\label{eq:R_definition}
\gamma_i=\underbrace{N_{i,1}^2I_{i,1}}_{M_{i,1}}\underbrace{N_{i,2}^2I_{i,2}}_{M_{i,2}},
\end{equation}
where $N_{i,j}, I_{i,j}$, and $M_{i,j}$, with $j\in \{1,2\}$, denote the multipath fading coefficients, the shadowing coefficients, and the resulting SNR of the two local scattering regions, respectively. {It is noted that due to the large distance between the two regions, and thus the existence of the keyhole effect, the SNR coefficients $M_{i,1}$ and $M_{i,2}$ are considered to be independent \cite{salo2006statistical}}. In this paper, the generic Nakagami-$m$ distribution has been assumed as a multipath fading model. In this context, the PDF of $N_{i,j}^2$ is given by \cite[eq. (2.20)]{B:Sim_Alou_Book}
\begin{equation}\label{eq:Nakagami_pdf}
f_{N_{i,j}^2}(x)=\frac{m_j^{m_j} x^{m_j-1}}{\Omega^{m_j}\Gamma(m_j)} \exp\left( -\frac{m_jx}{\Omega}\right),
\end{equation}
where $m_j$ is distribution's shaping parameter, related to the severity of the fading, i.e., as $m_j$ increases LoS conditions are approximated, $\Omega$ denotes the mean square value, and $\Gamma(\cdot)$ is the gamma function \cite[eq. (8.310/1)]{B:Ryzhik_book}. It should be noted that a one-to-one mapping exists between the $m_j$ parameter and the Rician $K$ factor, allowing the Nakagami-$m$ distribution to approximate the Rician distribution, which is a well-established channel model for LoS propagation environments \cite{B:Sim_Alou_Book}. 

The shadowing random fluctuations are modeled by the IG distribution with PDF given by
\begin{equation}\label{eq:pdf_igamma}
f_{I_{i,j}}(y)=\frac{\overline{\gamma}_{j}^{\alpha_{j}}}{\Gamma(\alpha_{j}) y^{\alpha_{j}+1}} \exp\left( -\frac{\overline{\gamma}_{j}}{y}\right),
\end{equation}
where $\alpha_{j}>1$ is the shaping parameter of the distribution, related to the severity of the shadowing, {i.e., lower values of $\alpha_j$ result in lighter shadowing conditions,} and $\overline{\gamma}_{j}$ denotes the scaling parameter. In order to evaluate the PDF of $\gamma_i$, the following two random-variables (RVs) are introduced
\begin{equation}\label{eq:definition_N_I}
\begin{split}
N_i&=\left(N_{i,1}N_{i,2}\right)^2\\
I_i&=I_{i,1}I_{i,2}.
\end{split}
\end{equation}
It should be noted that $N_i$ represents the square of a double-Nakagami-$m$ RV with PDF given as \cite[eq. (11)]{bithas2018v2v}
\begin{equation}\label{eq:DN_PDF}
f_{N_i}(y)=\frac{G \substack{2,0
\\ 0,2} \left( \frac{m_1m_2y}{\Omega^2}
\Big| \substack{-
\\ \null
\\ m_{1},m_{2}}\right)}{y\Gamma(m_{1})\Gamma(m_{2})},
\end{equation}
where $G \substack{m,n \\ p,q} \left[ \cdot | \cdot
\right]$ denotes the Meijer's G-function \cite[eq.
(9.301)]{B:Ryzhik_book}, which is a built-in function in many mathematical software packages, e.g., Mathematica, Maple, and thus it can be directly evaluated. In \eqref{eq:definition_N_I}, the RV $I_i$ models the double-IG (DIG) RV with PDF given by 
\cite{WCNC2019}
\begin{equation}\label{eq:DG_PDF}
\begin{split}
f_{I_i}(y)=\frac{2\overline{\gamma}^{\frac{a_1+a_2}2}}{\Gamma(a_1)\Gamma(a_2)}  
y^{-\frac{a_1+a_2}2-1} K_{a_2-a_1}\left(2 \left(\frac{\overline{\gamma}}{y}\right)^{1/2} \right),
\end{split}
\end{equation}
where $K_{v}(\cdot)$ denotes the modified Bessel function of the second kind and order $v$ \cite[eq. (8.407)]{B:Ryzhik_book}, while $\overline{\gamma}$ denotes the average SNR. By substituting \eqref{eq:DN_PDF} and \eqref{eq:DG_PDF} in the expression of the total probability theorem, i.e., \cite[eq. (4.80)]{B:Papoulis_book}, and using \cite[eqs. (14) and (21)]{C:Adamchik_Meijer}, the PDF of $\gamma_i$ can be deduced as
\begin{equation}\label{eq:PDF_final}
\begin{split}
f_{\gamma_i}&(\gamma)= \gamma^{-1} \mathcal{S}_1  G \substack{2,2
\\ 2,2} \left( \frac{m_1m_2\gamma}{\overline{\gamma}}
\Big| \substack{1-\alpha_2,1-\alpha_1
\\ \null
\\ m_1,m_2} \right)\\ 
 &\stackrel{(a)}{=} \Gamma(m_1+\alpha_2)\Gamma(m_2+\alpha_2)\Gamma(m_1+\alpha_1)\Gamma(m_2+\alpha_1)\mathcal{S}_1 \\ & \times \null_2\tilde{F}_1 \left( m_1+a_2,m_2+a_2;a_1+a_2+m_1+m_2; 1-\frac{1}{\tilde{\gamma}\gamma}\right)\gamma^{-a_2-1} \tilde{\gamma}^{-\alpha_2},
\end{split}
\end{equation}
where $\null_2\tilde{F}_1 \left( \cdot,\cdot;\cdot;\cdot\right)$ denotes the regularized Gauss hypergeometric function \cite[eq. (07.31.02.0001.01)]{E:wolfram}, $\mathcal{S}_1=\frac{1}{\Gamma(m_1)\Gamma(m_2)\Gamma(\alpha_1)\Gamma(\alpha_2)}$, $\tilde{\gamma}=\frac{m_1m_2}{\overline{\gamma}}$, while $(a)$ holds due to \cite[eq. (07.34.03.0873.01)]{E:wolfram}. Here, it should be noted that the mean value of $\gamma_i$ is given by 
\begin{equation}
    \mathbb{E}\left<\gamma_i\right>=\frac{\overline{\gamma}}{(\alpha_1-1)(\alpha_2-1)},
\end{equation}
where $\mathbb{E}\left<\cdot\right>$ denotes expectation. It is noteworthy that the mean value of $\gamma_i$ is independent from the (multipath) fading shaping parameters. {It is also noted that as $\alpha_i \rightarrow 1$ and $m_1 \rightarrow \infty$, the DSc model approximates single scattering composite fading model.}

\subsection{Single-Shadowing Communication Scenario}
In several cases, due to the geometrical characteristics of the propagation environment, it is expected that shadowing is present in just one of the two scattering regions. {This could be the scenario where a strong LoS or strong multipath components exist in only one of the two scattering regions, whereas the other one is suffering from severe shadowing phenomena and thus NLoS conditions. As a consequence, randomly vary mean values of the received envelopes can be found in only one of the two regions}. Under this single-shadowing (SS) assumption, \eqref{eq:R_definition} can be expressed as
\begin{equation}\label{eq:R_2nd_definition}
{\gamma}_i^s=\left(N_1N_2\right)^2 I_i.
\end{equation}
For evaluating the PDF of ${\gamma}_i^s$, \eqref{eq:pdf_igamma} and \eqref{eq:DN_PDF} are substituted in \cite[eq. (4.80)]{B:Papoulis_book}, and by using a similar approach as the one employed for the derivation of \eqref{eq:PDF_final}, the following simplified expression has been extracted
\begin{equation}\label{eq:PDF_SS}
\begin{split}
f_{{\gamma}_i^s} (\gamma) &= \mathcal{S}_2 \gamma^{-1} G \substack{2,1
\\ 1,2} \left( \tilde{\gamma} \gamma 
\Big| \substack{1-\alpha
\\ \null
\\ m_1,m_2} \right)\\  
 & \stackrel{(a)}{=} \Gamma(\alpha+m_1)\Gamma(\alpha+m_2) \mathcal{S}_2  \tilde{\gamma}^{m_1} \gamma^{m_1-1} U\left( a+m_1;m_1-m_2;\tilde{\gamma}\gamma\right),
\end{split}
\end{equation}
where $\alpha$ denotes the shaping parameter of the IG distribution, $U\left( \cdot;\cdot;\cdot\right)$ denotes the confluent hypergeometric function \cite[eq. (9.210/2) ]{B:Ryzhik_book}, and $\mathcal{S}_2=\frac{1}{\Gamma(m_1)\Gamma(m_2)\Gamma(\alpha)}$. It is noted that $(a)$ holds due to \cite[eq. (07.34.03.0719.01)]{E:wolfram}. The mean value of $\gamma_i$ is given as $\mathbb{E}\left<\gamma_i\right>=\frac{\overline{\gamma}}{(\alpha-1)}$. 

\section{Single UAV Communication Scenario}\label{sec:performance}
In this section, the performance of a communication system operating in a shadowed DSc fading environment has been analytically investigated. Moreover, selected numerical evaluated results are also presented.

\subsection{Outage Probability}
The OP is defined as the probability that the output SNR is below a predefined threshold $\gamma_T$, irrespective of the modulation scheme adopted. From the mathematical point of view, the OP is defined as
\begin{equation}\label{eq:OP_definition}
    P_{\rm out}= {\rm Pr} \left\{ \gamma_i \leq \gamma_T\right\}=F_{\gamma_i}(\gamma_T)=\int_0^{\gamma_T} f_{\gamma_i}(\gamma)d\gamma,
\end{equation}
where $F_{\gamma_i}(\gamma_T)$ denotes the CDF of $\gamma_i$ and ${\rm Pr} \left\{\cdot\right\}$ denotes the probability. For the DS scenario, substituting \eqref{eq:PDF_final} in \eqref{eq:OP_definition} and using \cite[eq. (26)]{C:Adamchik_Meijer}, results in the following closed-form expression for the CDF of $\gamma_i$
\begin{equation}\label{eq:CDF_DS}
\begin{split}
F_{\gamma_i}(\gamma)&= \mathcal{S}_1 G \substack{2,3
\\ 3,3} \left( \tilde{\gamma}\gamma
\Big| \substack{1-\alpha_2,1-\alpha_1,1
\\ \null
\\ m_1,m_2,0} \right).
\end{split}
\end{equation}
For the SS scenario, by employing \eqref{eq:PDF_SS}, \eqref{eq:CDF_DS} simplifies to 
\begin{equation}\label{eq:CDF_SS}
\begin{split}
F_{{\gamma}_i^s} (\gamma) &= \mathcal{S}_2  G \substack{2,2
\\ 2,3} \left( \tilde{\gamma}\gamma
\Big| \substack{1-\alpha,1
\\ \null
\\ m_1,m_2,0} \right).
\end{split}
\end{equation}

\subsection{Average Bit Error Probability}
For evaluating the BEP, the CDF-based approach will be adopted \cite{chen2004distribution}. Assuming binary phase shift keying (BPSK) modulation scheme, the BEP can directly be evaluated using the following integral 
\begin{equation}\label{eq:BEP_definition}
    P_{\rm be}=\frac{1}{2\sqrt{\pi}}\int_0^\infty F_{\gamma_i} (\gamma) \exp\left( -\gamma\right)d\gamma .
\end{equation}
Assuming DS communication environment, \eqref{eq:CDF_DS} is substituted in \eqref{eq:BEP_definition} and with the aid of \cite[eq. (21)]{C:Adamchik_Meijer}, the BEP of BPSK can be evaluated as
\begin{equation}\label{eq:SEP_DS}
\begin{split}
    P_{\rm be} &= \mathcal{S}_1 \left(2\sqrt{\pi}\right)^{-1}  G \substack{2,4
\\ 4,3} \left( \tilde{\gamma}
\Big| \substack{1-\alpha_2,1-\alpha_1,1,1/2
\\ \null
\\ m_1,m_2,0} \right).
\end{split}
\end{equation}
Moreover, for the SS scenario and by employing \eqref{eq:CDF_SS}, \eqref{eq:SEP_DS} simplifies to 
\begin{equation}\label{eq:SEP_SS}
\begin{split}
P_{\rm be} &= \mathcal{S}_2 \left(2\sqrt{\pi}\right)^{-1}  G \substack{2,3
\\ 3,3} \left( \tilde{\gamma}
\Big| \substack{1-\alpha,1,1/2
\\ \null
\\ m_1,m_2,0} \right).
\end{split}
\end{equation}

\subsection{Ergodic Capacity}
The ergodic channel capacity quantifies the maximum rate of reliable communications that can be supported by the subjected channel. Its definition is given by
\begin{equation}\label{eq:capacity_def}
    \begin{split}
        \bar{C} &= BW\int_0^\infty
 \log_2 (1+\gamma)f_{\gamma_i}(\gamma)d\gamma,    
\end{split}
\end{equation}
where $BW$ denotes the signal's transmission bandwidth. Assuming DS communication environment, \eqref{eq:PDF_final} is substituted in \eqref{eq:capacity_def} and with the aid of \cite[eq. (21)]{C:Adamchik_Meijer}, the capacity can be evaluated as
\begin{equation}\label{eq:DS_capacity}
    \begin{split}
        \bar{C} &=  BW \mathcal{S}_1 \ln \left( 2\right)^{-1}G \substack{4,3
\\ 4,4} \left(\tilde{\gamma}
\Big| \substack{1-\alpha_2,1-\alpha_1,0,1
\\ \null
\\ m_1,m_2,0,0} \right).
    \end{split}
\end{equation}
Moreover, for the SS scenario and by employing \eqref{eq:PDF_SS}, \eqref{eq:DS_capacity} simplifies to 
\begin{equation}\label{eq:capacity_SS}
\begin{split}
\bar{C} &= BW \mathcal{S}_2 \ln \left( 2\right)^{-1}  G \substack{4,2
\\ 3,4} \left(\tilde{\gamma}
\Big| \substack{1-\alpha,0,1
\\ \null
\\ m_1,m_2,0,0} \right).
\end{split}
\end{equation}

\subsection{Numerical Results}
\begin{figure}[t]
\centering
\includegraphics[keepaspectratio,width=4in]{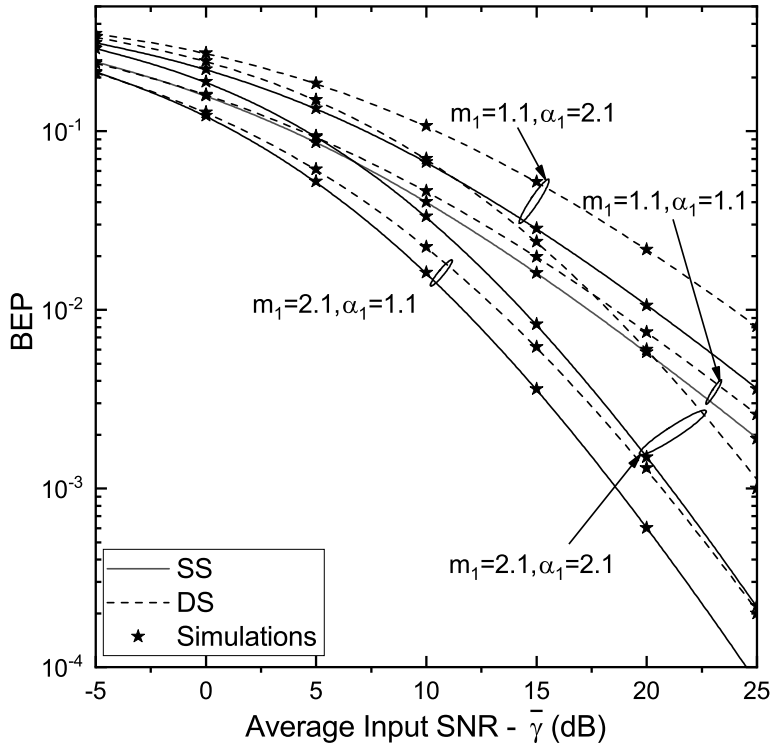}
\caption{BEP as a function of the average input SNR.}\label{fig:1}
\end{figure}

\begin{figure}[t]
\centering
\includegraphics[keepaspectratio,width=4in]{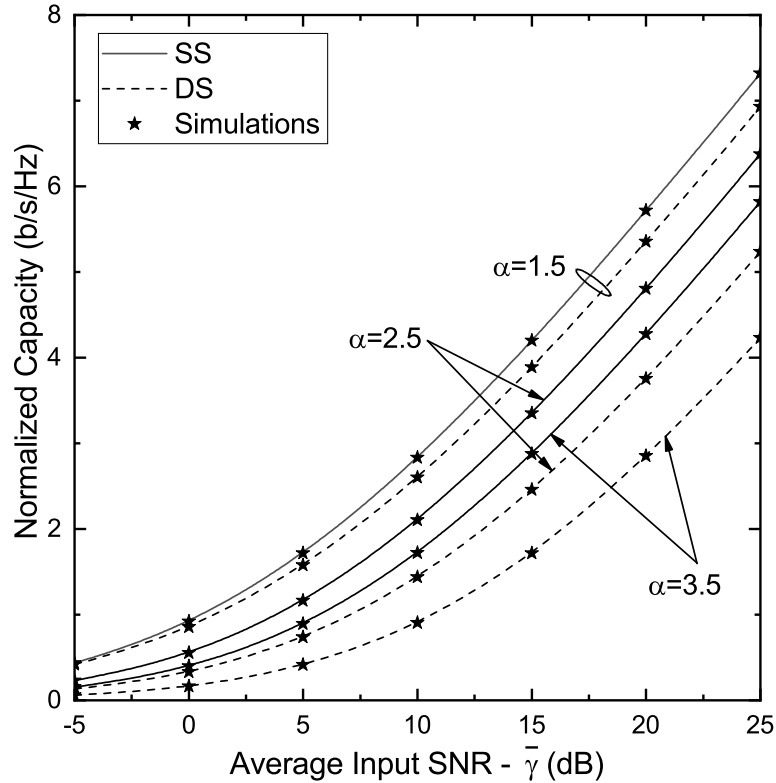}
\caption{Normalized capacity as a function of the average input SNR.}\label{fig:2}
\end{figure}

Next, using the stochastic analysis developed previously, various numerically evaluated results are illustrated. In Fig.~\ref{fig:1}, the BEP of BPSK modulation scheme has been plotted as a function of the average input SNR $\overline{\gamma}$, for both DS and SS channel models. For obtaining this figure, \eqref{eq:SEP_DS} and \eqref{eq:SEP_SS} have been employed, while $m_2=m_1+0.3$ and $\alpha_2=\alpha_1+0.3$. It is shown that the performance improves when fading parameter $m_i$ increases and/or shadowing parameter $\alpha_i$ decreases, with the former having highest impact. Moreover, it is interesting to note that an important difference between the performance of DS and SS channels is observed. This difference can easily be explained if one considers that in DS scenario, shadowing is present in both scattering regions around the Tx and the Rx. On the other hand, in the SS scenario, shadowing is present in only one of the two regions, which results in improved communication conditions. In Fig.~\ref{fig:2}, the normalized capacity $C/BW$ is plotted as a function of $\overline{\gamma}$ for different values of $\alpha_i$. To obtain this figure, \eqref{eq:DS_capacity} and \eqref{eq:capacity_SS} have been employed, with $m_1=1.5$, $m_2=1.8$, and $\alpha_2=\alpha_1+0.1$. In this figure, it is interesting to note that the performance decreases as $\alpha_i$s increase, with a reducing rate. Finally, {Monte Carlo simulation results}\footnote{Values of the RVs under investigation have been generated in MATLAB\texttrademark using their definitions, i.e., \eqref{eq:R_definition} and \eqref{eq:R_2nd_definition} for the DS and SS scenarios, respectively.} have been also included in Figs.~\ref{fig:1}-\ref{fig:2}, depicting in all cases the correctness of the proposed framework.

\section{UAV Selection Strategy}\label{sec:UAV_selection}
In scenarios where $L$ UAVs operate as flying access points, the users' quality of experience might be improved by exploiting the additional available resources. In this context, a quite efficient and practical approach, which is also proposed in this paper, is to select the UAV access point to be connected to, according to a specific UAV association policy. A well-known criterion that has been widely used in similar scenarios is to select the UAV that offers the best instantaneous received SNR per coherence time (CT) interval, e.g., \cite{Ji2019}. Apparently, this approach is mainly based on the assumption of  accurate estimations of the received SNR values. However, these estimates might be inaccurate especially in highly dynamic communication environments and/or in cases with increased distance between the Tx and the Rx. It is safe to say that both these scenarios characterize the UAV-enabled communications. An alternative approach, which is also adopted in this paper, is to select the associated UAV by exploiting the information that is available for the shadowing behavior \cite{6310073,8584097}. This approach is capitalizing on the larger decorrelation distance of the large-scale fading as compared to the small-scale fading CT. In this context, the selected UAV is the one providing the maximum averaged received power, i.e., shadowing variable, over a predetermined time interval. It is noted that this approach may be applied in both the uplink and downlink, since in the latter case an index is fed back to the Tx in order to proceed with the signal transmission. 

With the above in mind, the received SNR of the proposed shadowing-based UAV-selection mechanism is given by
\begin{equation}
\gamma_{\rm out}=N_{i} \max_{\substack{1\leq i\leq L }} \{I_{i}\}=N_{i} I_{\max},
\end{equation}
where $N_{i}$ depends on the multipath fading of the selected UAV. In order to analytically evaluate the statistics of $\gamma_{\rm out}$, the PDF of $I_{\rm max}$ should be evaluated.

\subsection{Double-Shadowing Scenario}\label{subsec:DS_scenario}
As far as the DS scenario is concerned, the PDF of $I_{\max}$ can be evaluated with the aid of the following theorem.
\begin{theorem}
The PDF of $I_{\max}=\max\limits_{\substack{1\leq i\leq L }} \{I_{i}\}$ is given as
\begin{equation}\label{eq:PDF_max}
    \begin{split}
     f_{I_{\max}}(\gamma) = &   \frac{2L}{\Gamma(\alpha_1)\Gamma(\alpha_2)}\sum_{i_1=0}^{L-1} \sum_{i_2=0}^{i_1}
     \binom{L-1}{i_1}\binom{i_1}{i_2} \sum_{\substack{n_0, \cdots, n_{2\alpha_1-1}=0\\ n_0+ \cdots+ n_{2\alpha_1-1}=i_2}}^{i_2} \mathcal{A} \frac{\overline{\gamma}^{q/2}}{\gamma^{q/2+1}} \\ & \times  \exp\left( -2 i_2 \left(\frac{\overline{\gamma}}{\gamma}\right)^{1/2} \right) K_{\alpha_2-\alpha_1} \left( 2\left(\frac{\overline{\gamma}}{\gamma}\right)^{1/2}\right),
    \end{split}
\end{equation}
where $$\sum_{\substack{n_0, \cdots, n_{2\alpha_1-1}=0\\ n_0+ \cdots+ n_{2\alpha_1-1}=i_2}}^{i_2} = \underset{n_0+n_1+\cdots+n_{2\alpha_1-1}=i_2}{\sum_{n_0=0}^{i_2} \sum_{n_1=0}^{i_2}\cdots \sum_{n_{2\alpha_{1}-1}=0}^{i_2}},   $$
while $\mathcal{A}=\frac{i_2!}{n_0! n_1! \cdots n_{2a_1-1}!} \prod_{j=1}^{2a_1-1} \left( \frac{2^j}{j!}\right)^{n_{j+1}} (-1)^{i_1+i_2},$ and $q=a_1+a_2+\sum_{j=1}^{2a_1-1}j n_{j+1}$.
\end{theorem}
\begin{IEEEproof}
See Appendix~\ref{App:1}.
\end{IEEEproof}
Since $\gamma_{\rm out}$ is actually the product of two independent RVs, its PDF and CDF expressions can, respectively, be defined as
\begin{equation}\label{eq:final_PDF_CDF_defs}
    \begin{split}
        f_{\gamma_{\rm out}} (\gamma) & = \int_0^\infty \frac{1}{x} f_{N_i} \left( \frac{\gamma}{x}\right) f_{I_{\max}}(x)dx\\
        F_{\gamma_{\rm out}} (\gamma) & = \int_0^\infty  F_{N_i} \left( \frac{\gamma}{x}\right) f_{I_{\max}}(x)dx.
    \end{split}
\end{equation}
The PDF of $\gamma_{\rm out}$ can be derived by substituting \eqref{eq:DN_PDF} and \eqref{eq:PDF_max} in \eqref{eq:final_PDF_CDF_defs} and using \cite[eq. (21)]{C:Adamchik_Meijer} yielding to 
\begin{equation}\label{eq:PDF_DS}
    \begin{split}
     f_{\gamma_{\rm out}}(\gamma) = &   \sum_{i_1=0}^{L-1} \sum_{i_2=0}^{i_1}
       \binom{L-1}{i_1}\binom{i_1}{i_2} \sum_{\substack{n_0, \cdots, n_{2\alpha_1-1}=0\\ n_0+ \cdots+ n_{2\alpha_1-1}=i_2}}^{i_2}   \\ &  \frac{\times \gamma^{-1}\mathcal{A}L \mathcal{S}_1}{\left(i_2+1 \right)^{q-1/2}} G \substack{2,2
\\ 2,2} \left( \frac{\tilde{\gamma}\gamma}{ (i_2+1)^2}
\Big| \substack{\frac{3/2-q}{2},\frac{5/2-q}{2}
\\ \null
\\ m_1,m_2} \right).
    \end{split}
\end{equation}
Moreover, the corresponding CDF is given by
\begin{equation}\label{eq:CDF_DS_out}
    \begin{split}
     F_{\gamma_{\rm out}}(\gamma) =  &   \sum_{i_1=0}^{L-1} \sum_{i_2=0}^{i_1}
      \binom{L-1}{i_1}\binom{i_1}{i_2} \sum_{\substack{n_0, \cdots, n_{2\alpha_1-1}=0\\ n_0+ \cdots+ n_{2\alpha_1-1}=i_2}}^{i_2}   \\ & \times \frac{\mathcal{A}L\mathcal{S}_1}{\left(i_2+1 \right)^{q-1/2}}  G \substack{2,3
\\ 3,3} \left( \frac{\tilde{\gamma}\gamma}{ (i_2+1)^2}
\Big| \substack{1,\frac{3/2-q}{2},\frac{5/2-q}{2}
\\ \null
\\ m_1,m_2,0} \right).
    \end{split}
\end{equation}
For evaluating the ASNR, \eqref{eq:PDF_DS} is substituted in the corresponding definition, i.e.,
\begin{equation}\label{eq:ASNR_def}
    \begin{split}
        \bar{\gamma}_{\rm out} &= \int_0^\infty
 \gamma f_{\gamma_{\rm out}}(\gamma)d\gamma,    
\end{split}
\end{equation}
and using \cite[eq. (7.811/4)]{B:Ryzhik_book}, the following closed-form expression is deduced
\begin{equation}\label{eq:moments_DS}
    \begin{split}
    \overline{\gamma}_{\rm out} = &   \frac{L}{\Gamma(\alpha_1)\Gamma(\alpha_2)}
      \sum_{i_1=0}^{L-1} \sum_{i_2=0}^{i_1} \binom{L-1}{i_1}\binom{i_1}{i_2}
      \\   & \times \sum_{\substack{n_0, \cdots, n_{2\alpha_1-1}=0\\ n_0+ \cdots+ n_{2\alpha_1-1}=i_2}}^{i_2} \frac{\mathcal{A}\overline{\gamma} 2^{7/2-\alpha_1-\alpha_2} \sqrt{\pi} \Gamma\left(q-5/2 \right)}{\left(i_2+1 \right)^{q-5/2}}  .
    \end{split}
\end{equation}
An important insight that stems out from \eqref{eq:moments_DS} is that the shaping parameters of the double-Nakagami distribution, i.e., the behavior of the multipath fading, have no impact to the ASNR of the proposed  system.

\subsection{Single-Shadowing Scenario}
As far as the SS scenario is concerned, the PDF of $I_{\max}$ is given in \cite[eq. (15)]{8584097}.  Based on this expression, and following a similar approach as the one presented in Section ~\ref{subsec:DS_scenario}, the following closed-form expression for the PDF of $\gamma_{\rm out}$ is deduced
\begin{equation}
    \begin{split}
       f_{\gamma_{\rm out}} (\gamma)& = \frac{\mathcal{B} \mathcal{S}_2}{\gamma}\sum_{\substack{n_1, \cdots, n_{\alpha}=0\\ n_1+ \cdots+ n_{\alpha}=L-1}}^{L-1}   G \substack{2,1
\\ 1,2} \left( \frac{\tilde{\gamma}\gamma}{L}
\Big| \substack{1-p
\\ \null
\\ m_1,m_2} \right),
    \end{split}
\end{equation}
where $$\mathcal{B}=L^{-p} \frac{L!}{n_1!n_2!\cdots n_\alpha!}\prod_{j=1}^{\alpha} \left( \frac{1}{(j-1)!}\right)^{n_j},$$ and $p=\alpha+\sum_{j=1}^{\alpha} (j-1)n_j$. Moreover, the corresponding expression for the CDF of $\gamma_{\rm out}$ is given by
\begin{equation}\label{eq:CDF_SS_out}
    \begin{split}
       F_{\gamma_{\rm out}} (\gamma) & = \mathcal{B} \mathcal{S}_2\sum_{\substack{n_1, \cdots, n_{\alpha}=0\\ n_1+ \cdots+ n_{\alpha}=L-1}}^{L-1}    G \substack{2,2
\\ 2,3} \left( \frac{\tilde{\gamma}\gamma}{ L}
\Big| \substack{1,1-p
\\ \null
\\ m_1,m_2,0} \right).
    \end{split}
\end{equation}
Finally, for the scenario of SS, the ASNR is given by
\begin{equation}\label{eq:moments_SS}
    \begin{split}
       \overline{\gamma}_{\rm out}  & = \frac{L\overline{\gamma}}{\Gamma(\alpha)}\sum_{\substack{n_1, \cdots, n_{\alpha}=0\\ n_1+ \cdots+ n_{\alpha}=L-1}}^{L-1}\mathcal{B}   \Gamma\left( p-1\right).
    \end{split}
\end{equation}

Here, it should be noted that the analytical results presented in the previous sections allow the system designer to predict the system performance in various communication scenarios, where this particular channel model can be applied. In the next section, it will be shown that the proposed channel model provides an excellent fit to real A2G communication environments.

\subsection{Numerical Results}
\begin{figure}[t]
\centering
\includegraphics[keepaspectratio,width=4in]{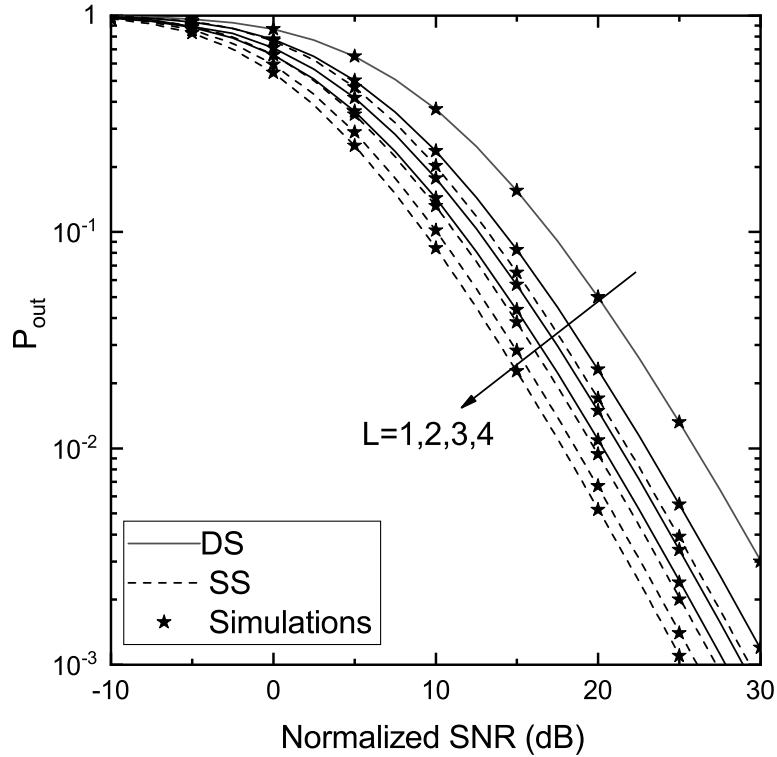}
\caption{OP as a function of the normalized outage threshold for different architectures.}\label{fig:3}
\end{figure}

\begin{figure}[t]
\centering
\includegraphics[keepaspectratio,width=4in]{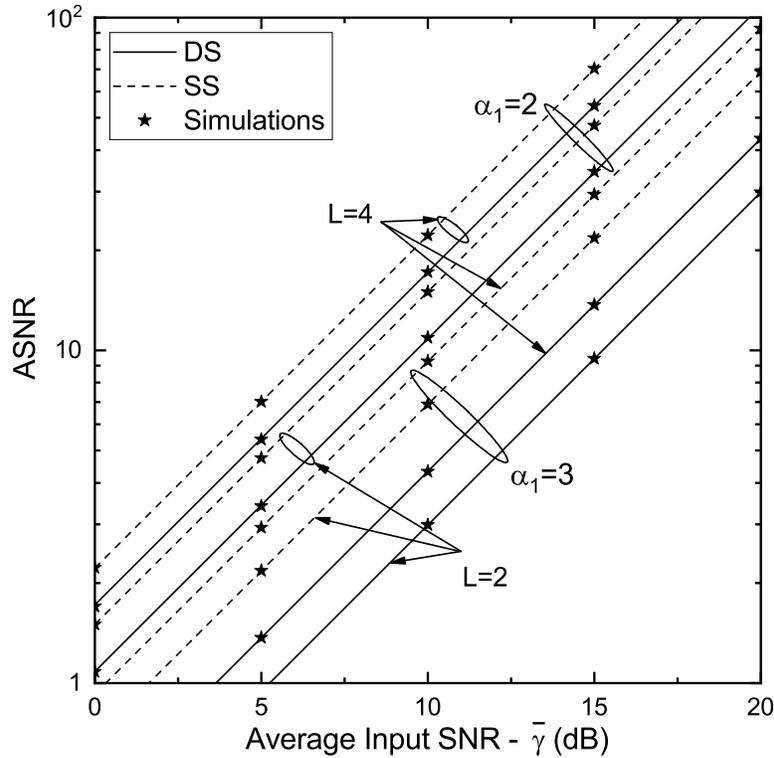}
\caption{ASNR as a function of the average input SNR for different architectures.}\label{fig:4}
\end{figure}

In Fig.~\ref{fig:3}, assuming the proposed UAV-selection strategy, the OP is plotted as a function of the normalized outage threshold ($\overline{\gamma}/\gamma_T$), for both channel models and different values {of the number of UAVs} $L$. For obtaining this figure, \eqref{eq:CDF_DS_out} and \eqref{eq:CDF_SS_out} have been employed, with $m_1=1.5,$ $m_2=1.8,$ $\alpha_1=2,$ $\alpha_2=2.5$. In this figure, it is depicted that the performance improves as the number of UAVs increases. Moreover, the performance gain degrades as $L$ increases. Similar outcomes can be extracted in Fig.~\ref{fig:4}, where the ASNR is plotted as a function of $\overline{\gamma}$ also for both channel models and different values of $\alpha_i$. For obtaining this figure, \eqref{eq:moments_DS} and \eqref{eq:moments_SS} have been employed, with $\alpha_2=\alpha_1+0.5$. Once more, it is shown that the performance is better in DS model and as $\alpha_i$s increase. It is worth noting that the gap among the curves is higher when severe shadowing conditions exist, i.e., $\alpha_1=3$, as compared to the mild shadowing scenarios. 

\section{Empirical Results and Comparisons}
In this section, empirical results obtained in a measurement campaign will be exploited. These results will be used to justify the good fit of the considered channel model to empirical data as well as to prove the performance improvement achieved by the proposed UAV-selection strategy in real-world scenarios. The measurement campaign took place in Prague, in an urban pedestrian environment, with the Rx positioned on the street level in the middle of a crossroad, while the Tx was mounted on the bottom part of a zeppelin-type airship. The airship moved at a constant speed of 6.2 m/s, in a pre-defined route over the position where the Rx was located. The height above the ground was maintained at approximately 200 m while the Rx was kept stationary, mounted on a tripod at 1.7 m above the ground. The airship was equipped with a GPS sensor, enabling the Tx to point the antennas constantly towards the Rx by using a positioner attached to the bottom part of the ship and allowed the distance between the Rx and the Tx to be calculated. The transmitted signals were received by a single dual-linearly polarized rectangular patch antenna (collocated). More details for the measurement campaign can be found in \cite{nikolaidis2017dual}.
\begin{table}[]\label{tab:1}
\renewcommand{\arraystretch}{1.4}
\caption{K-L and K-S Goodness-of-Fit.}\label{tab:1} \centering
\begin{tabular}{|l|l|l|}
\hline
\textbf{Test}               & K-L     & K-S     \\ \hline
$\bf dN_I$ \textbf{(DIG)}   & 6.7\%   & 85.96\% \\ \hline
$\bf dN_I$ \textbf{(SIG)}   & 6.75\%  & 71.44\% \\ \hline
$\bf dK_G$                  & 7.24\%  & 42.62\% \\ \hline
\end{tabular}
\end{table}

Next, based on the statistical framework developed on the previous section and the empirical data collected in \cite{nikolaidis2017dual}, various comparative results are presented. More specifically, three composite DSc distributions have been investigated, namely the double-Nakagami double-gamma, which is widely known as double-generalized K ($\rm dk_G$) \cite{5426028}, the proposed double-Nakagami double IG (denoted as $dN_I$ (DIG)), and the double-Nakagami with single IG (denoted as $dN_I$ (SIG)). It is noted that the method of moments was employed for estimating the parameters of all distributions. In this context, in order to investigate the suitability of the proposed distributions in fitting the empirical data sets, well-known goodness-of-fit  tests have been applied. In particular, the Kullback–Leibler (K-L) divergence criterion \cite{kullback1951information} and the Kolmogorov–Smirnov (K-S) test \cite[eq. (8.320]{B:Papoulis_book} were used. As far as the K-L distance is concerned, it is given by \cite{sen2008vehicle}
\begin{equation}
d_{KL}=\frac12 \left( \sum_i p_i \log \left( \frac{p_i}{q_i}\right)+\sum_i q_i \log \left( \frac{q_i}{p_i}\right)\right),
\end{equation}
where $p_i$ and $q_i$ are the sets of the simulated and empirical PDF values, respectively. The distribution that fits best to the measured data is the one that minimizes the K-L distance. Moreover, the K-S is defined as 
\begin{equation}
    d_{KS}=\underset{x}{\max} \left| F_e(x)-F_t(x)\right|,
\end{equation}
where $F_e(x)$ and $F_t(x)$ denote the empirical and theoretical CDFs, respectively. By assuming a $95\%$ confidence interval to all candidate distributions, a satisfactory fit can be ensured. The results are tabulated in Table~\ref{tab:1}. In this table, it is shown that the distance
K-L remains below $10\%$, with the models incorporating the IG distribution holding always the smallest values for this criterion. Moreover, the passing rates of K-S tests are quite encouraging in all cases, with the DIG distribution providing the best performance.

\begin{figure}[t]
\centering
\includegraphics[keepaspectratio,width=4in]{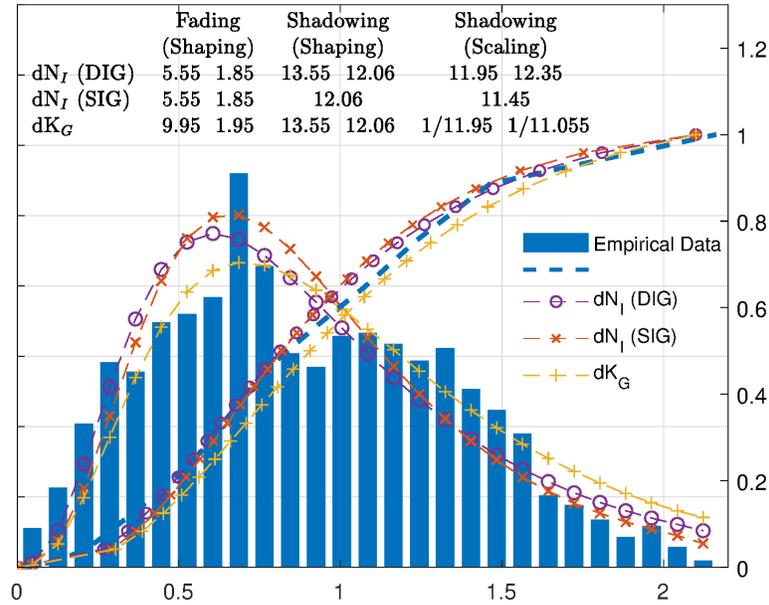}
\caption{Empirical and theoretical PDFs and CDF comparisons of a route segment.}\label{fig:SEG_PDF}
\end{figure}

The good fit can be also verified in Fig.~\ref{fig:SEG_PDF}, where plots with the theoretical and empirical PDFs and CDFs are provided. {The estimated values of the distributions' parameters, used for obtaining these two figures, are provided in the Table included in Fig.~\ref{fig:SEG_PDF}}. A general observation is that the investigated scenarios are characterized by NLoS conditions (for the end-to-end link) and strong scattering phenomena at the Rx side, due to the moving vehicles and surrounding buildings. This can be easily verified by the shapes of the PDFs and CDFs in Fig.~\ref{fig:SEG_PDF}, in which the direct LoS component is almost totally blocked. In all scenarios, the double-scattered signals may emanate from reflections along the vehicles and the multiple building facades from the structures surrounding the Rx.

\begin{figure}[t]
\centering
\includegraphics[keepaspectratio,width=4in]{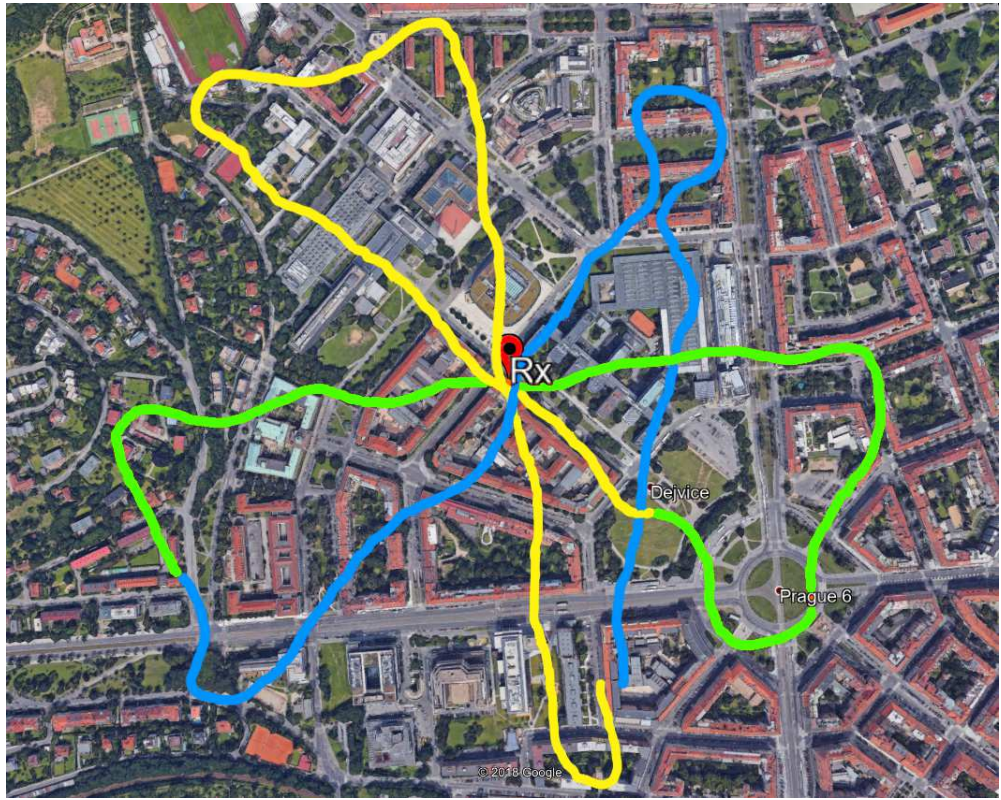}
\caption{Measurement environment and the airship route. The red pin indicates the position of the Rx.}\label{fig:RouteOfUAVs}
\end{figure}

\begin{figure}[t]
\centering
\includegraphics[keepaspectratio,width=4in]{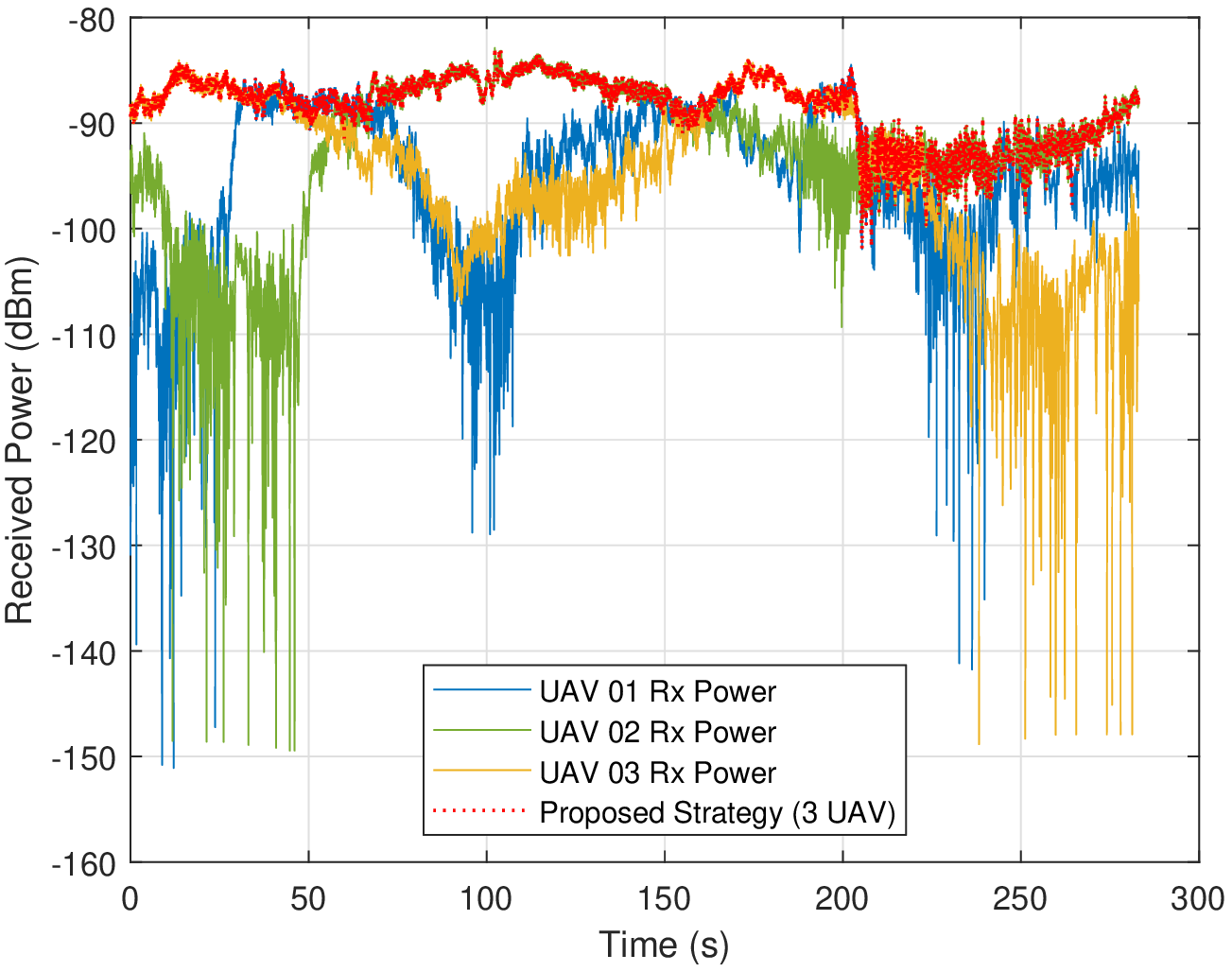}
\caption{Rx Power from 3 UAVs and the proposed selection strategy.}\label{fig:Rx_Power}
\end{figure}
\begin{figure}[t]
\centering
\includegraphics[keepaspectratio,width=4in]{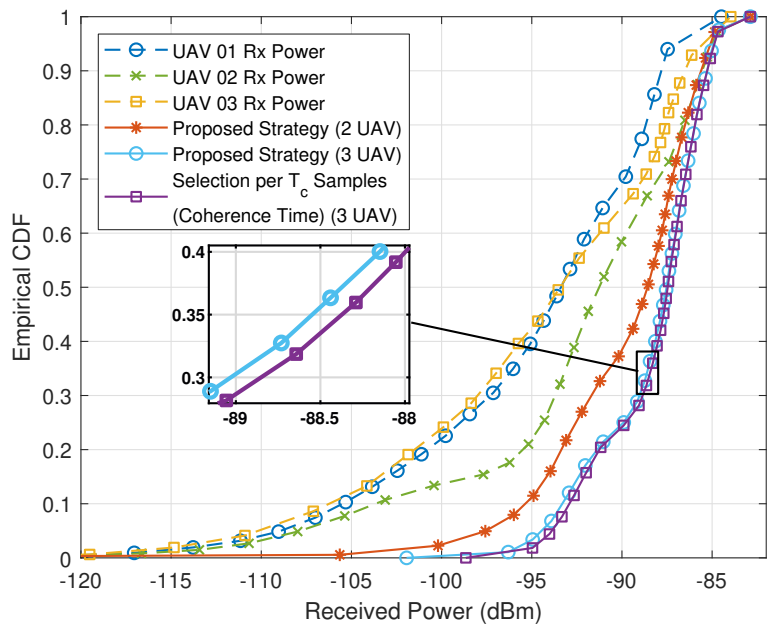}
\caption{Rx Power from all UAVs along with the proposed selection strategy and selection per sample.}\label{fig:Proposed_Strategy}
\end{figure}

The collected data, related to the received signal power, were split in $2$  (or $3$) segments, in order to emulate the received power from $2$  (or $3$) independent UAVs. Each of these segments has been assumed to belong to a ``virtual" UAV (indicated with specific colors in Figs.~\ref{fig:RouteOfUAVs}~and~\ref{fig:Rx_Power}, i.e., blue for UAV 1, green for UAV 2, and yellow for UAV 3). Fig.~\ref{fig:RouteOfUAVs} depicts the ``virtual" UAV's routes. In this context, we were able to compare the performance of the proposed UAV selection strategy with the corresponding of the ``3 independent UAVs". First of all, in Fig.~\ref{fig:Rx_Power}, the received power acquired from the 3 ``virtual" UAVs is depicted. In the same figure, the corresponding power of the output of the proposed scheme is also included, in which the Rx selects the UAV that offers the best SNR, on average, per stationarity region. This average SNR value is evaluated with the help of a sliding window of $\overline{W} = 40 \lambda$, where $\lambda=15$cm, which corresponds to 242 samples. In Fig.~\ref{fig:Proposed_Strategy}, the empirical CDF of the received power is plotted for the virtual UAVs, the proposed scheme, and an alternative UAV-selection approach. In the latter one, the UAV selection process is performed per channel's CT interval and is the most ordinary strategy that is frequent employed in scenarios where the best SNR principle is adopted. The performance improvement due to the proposed shadowing-based selection strategy is apparent from  Fig.~\ref{fig:Proposed_Strategy}. This observation holds also for the case where selection is performed between 2 UAVs. What is of particular interest to be noted is that the proposed strategy offers a comparable performance with the CT-based selection. However, for this communication system, the CT was equal to $0.7\lambda$ (or $170$ samples). Thus, it is quite interesting to note that comparing shadowing-based selection, performed per stationarity region, i.e., $40\lambda$, with CT selection, performed in every $0.7\lambda$, results in $57$ times less channel comparisons required for the proposed approach, without any important loss on the performance. {Therefore, this reduction on the channel comparisons is expected to reduce i) the signal processing complexity, ii) the switching operations (among the UAV links), and iii) the requirements for the feedback overhead.}

\section{Conclusions}
In this paper, a new {generic} shadowed DSc channel model has been proposed for describing the received signal behavior in UAV-enabled communication scenarios where DSc coexists with shadowing. Capitalizing on the {convenient mathematical form} of the new composite fading model, important stochastic metrics for the received SNR of a UAV communication system have been analytically derived. These expressions are then used to study the performance of the subjected system in terms of the OP, the BEP, and the channel capacity. Moreover, a new UAV-selection policy has been proposed that intends to offer improved performance with reduced {signal processing} complexity. This new scheme exploits the slow variations of the signal mean values for the UAV association. In this sense, novel closed-form expressions have been derived for the new scheme's statistics of the output SNR and then used to study its performance. The accuracy of the channel model as well as the applicability of the proposed selection strategy have been also verified by empirical data collected in an A2G measurement campaign, proving also the usefulness of the derived results to real-world scenarios. {In our future work, various interesting issues will be addressed, including the impact of interfering effects in a multi-UAV network as well as the adoption of appropriate interference mitigation techniques.}

\appendices
\renewcommand{\theequation}{A-\arabic{equation}}
\setcounter{equation}{0}
\section{Proof for Theorem 1}\label{App:1}
In this Appendix, the proof for Theorem 1 is provided. With the help of the order statistics of independent RVs, the PDF of $I_{ \max}$ is defined as
\begin{equation}\label{eq:Ap1}
    f_{I_{\max}} (\gamma)=Lf_{I_i}(\gamma)F_{I_i}(\gamma)^{L-1},
\end{equation}
where $f_{I_i}(\gamma)$ is given by \eqref{eq:DG_PDF} and $F_{I_i}(\gamma)$ is given by 
\begin{equation}\label{Ap:2}
\begin{split}
    F_{I_i}(\gamma)&= 1-\frac{1}{\Gamma(\alpha_1)\Gamma(\alpha_2)} G \substack{2,1
\\ 1,3} \left( \frac{\overline{\gamma}}{ \gamma}
\Big| \substack{1
\\ \null
\\ \alpha_1,\alpha_2,0} \right)\\
&\stackrel{(a)}{=} 1-\frac{2^{1-2\alpha_1}\sqrt{\pi}}{\Gamma(\alpha_1)\Gamma(\alpha_2)} \gamma\left( 2\alpha_1,2\sqrt{\frac{\overline{\gamma}}{\gamma}}\right)
\\
&\stackrel{(b)}{=} 1-\frac{2^{1-2\alpha_1}\sqrt{\pi} \Gamma(2\alpha_1)}{\Gamma(\alpha_1)\Gamma(\alpha_2)} \left[1-\exp\left( -2\left(\frac{\overline{\gamma}}{\gamma}\right)^{1/2}\right)\right]  \sum_{k=0}^{2\alpha_1-1} \frac{1}{k!} \left( 2\left(\frac{\overline{\gamma}}{\gamma}\right)^{1/2}\right)^k.
\end{split}
\end{equation}
In \eqref{Ap:2}, $(a)$ holds due to \cite[eq. (07.34.03.0732.01)]{E:wolfram} (with $\gamma\left(\cdot,\cdot\right)$ denoting the lower incomplete gamma function \cite[eq. (8.350/1)]{B:Ryzhik_book}), while $(b)$ holds due to \cite[eq. (8.353/1)]{B:Ryzhik_book}, under the assumption of $2\alpha_1\in \mathbb{N}$. Next, in order to provide a simplified expression for the $ F_{I_i}(\gamma)^{L-1}$ , the binomial identity has been employed twice resulting in 
\begin{equation}
    \begin{split}
         &F_{I_i}(\gamma)^{L-1} = \sum_{i_1=0}^{L} \sum_{i_2=0}^{i_1}\binom{L}{i_1} \binom{i_1}{i_2} (-1)^{i_1+i_2} \exp\left( - 2i_2\left(\frac{\overline{\gamma}}{\gamma}\right)^{1/2}\right) \left(\sum_{k=0}^{2\alpha_1-1} \frac{2^k}{k!}  \left(\frac{\overline{\gamma}}{\gamma}\right)^{k/2} \right)^{i_2} .
    \end{split}
\end{equation}
Finally, by employing the multinomial identity, \cite[eq. (24.1.2)]{abramowitz1964handbook}
\begin{equation}\begin{split}
    \left( x_1+x_2+\cdots x_L\right)^n&= \underset{n_1+n_2+\cdots+n_L=n}{\sum_{n_1=0}^{n} \sum_{n_2=0}^{n}\cdots \sum_{n_L=0}^{n}} \frac{n!x_1^{n_1}x_2^{n_2}\cdots x_L^{n_L}}{n_1!n_2!\cdots n_L!} 
\end{split}
\end{equation}
and using \eqref{eq:Ap1} and \cite[eq. (8.468)]{B:Ryzhik_book} finally yields \eqref{eq:PDF_max} and also completes this proof.

\section*{Acknowledgement}
The authors would like to acknowledge Prof. N. Hatzidiamantis for his valuable help on implementing the K-S tests. P. S. Bithas dedicates this article to Spilio G. Bitha.
\bibliographystyle{IEEEtran}
\bibliography{IEEEabrv,references}

\begin{IEEEbiography}[{\includegraphics[width=1in,height
=1.25in,clip,keepaspectratio]{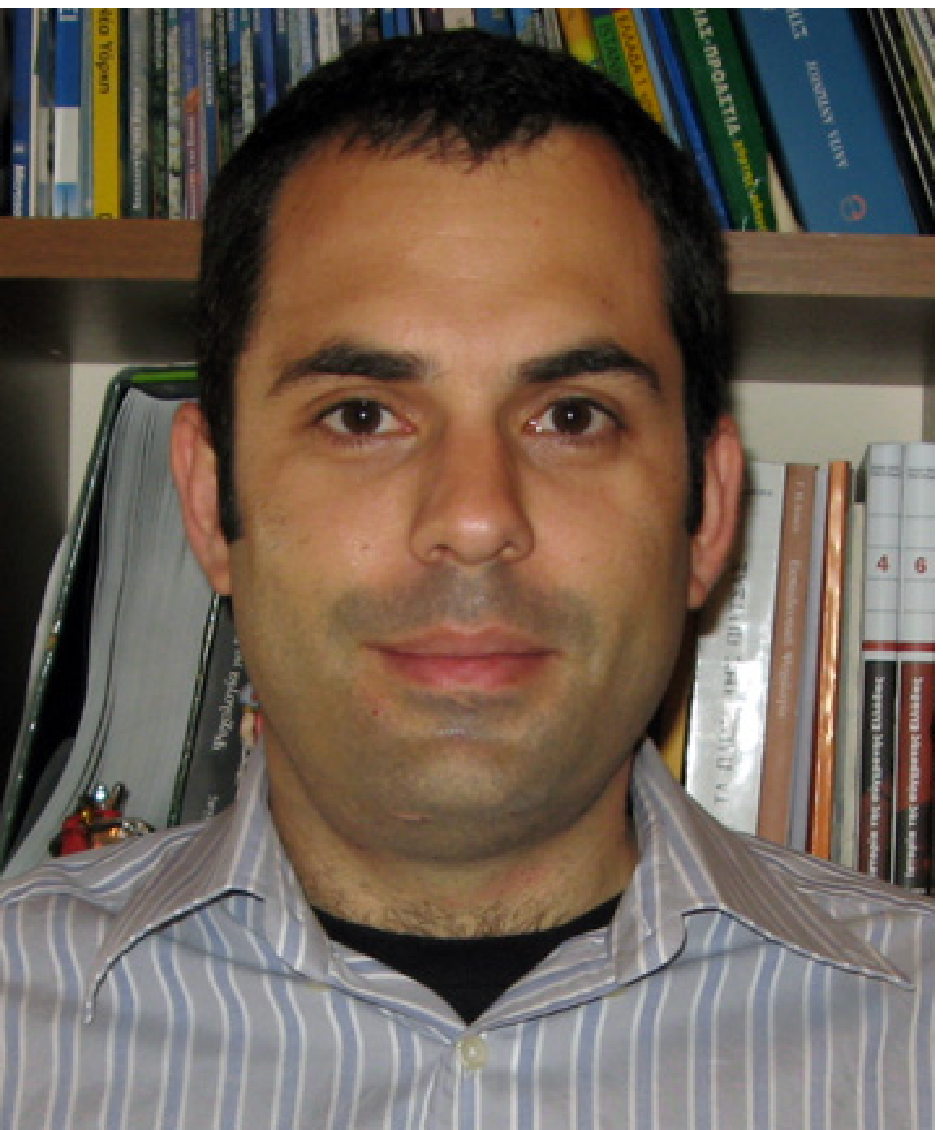}}]{Petros
S. Bithas}(S'04-M'09-SM'19) received the Diploma (5 years) in electrical and computer engineering and PhD degree, with specialization in ``Wireless Communication Systems", from the University of Patras, Greece, in 2003 and 2009, respectively. During 2009-2018 he was affiliated with the Department of Electronics Engineering of the Technological Educational Institute of Piraeus, Greece. Since the November of 2010, he is an associate researcher at the Department of Digital Systems, University of Piraeus (UNIPI), Greece, where he participates in a number of R\&D projects. He is currently an assistant Professor at the General Department of the National \& Kapodistrian University of Athens, Greece. Dr Bithas serves on the Editorial Board of the International Journal of Electronics and Communications (ELSEVIER) and Telecom (MDPI). He has been selected as an ``Exemplary Reviewer" of \textsc{IEEE Communications Letters} and \textsc{IEEE Transactions on Communications} in 2010 and 2020, respectively, while he is also co-recipient of Best Paper Awards at the IEEE International Symposium on Signal Processing and Information Technology, 2013. He has published 39 articles in International scientific journals and 39 articles in the proceedings of International conferences. His current research interests include stochastic modeling of wireless communication channels as well as design and performance analysis of vehicular communication systems.
\end{IEEEbiography}

\begin{IEEEbiography}[{\includegraphics[width=1in,height
=1.25in,clip,keepaspectratio]{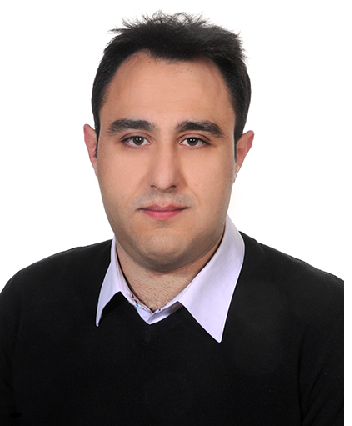}}]{Viktor Nikolaidis}(S’02) received the B.Sc. degree in technology education and digital systems from the UNIPI, Greece, in 2005, and the M.Sc. degree in mobile and satellite communications from the University of Surrey, U.K., in 2007. He is currently pursuing the Ph.D. degree with the Department of Digital Systems, UNIPI. From 2008 to 2009, he was with the European Space Agency, Hands-On Projects Section, Education Office, European Space Research and Technology Centre, The Netherlands. In 2009, he attended a Space Studies Program with the NASA Ames Research Center, Mountain View, CA, USA, organized by the International Space University, Strasbourg, France. In 2013, he was with the Telecommunication Systems Laboratory, UNIPI, as a Researcher. Since then, he has been involved in a number of research projects, including channel measurements and characterization for V2V and air-to-ground communication systems. His current research interests include MIMO techniques and channels, channel measurements, characterization and simulation for wireless, and satellite communication systems. Mr. Nikolaidis served as a Treasurer of the local IEEE Student Branch, UNIPI, from 2003 to 2006.
\end{IEEEbiography}

\begin{IEEEbiography}[{\includegraphics[width=1in,height
=1.25in,clip,keepaspectratio]{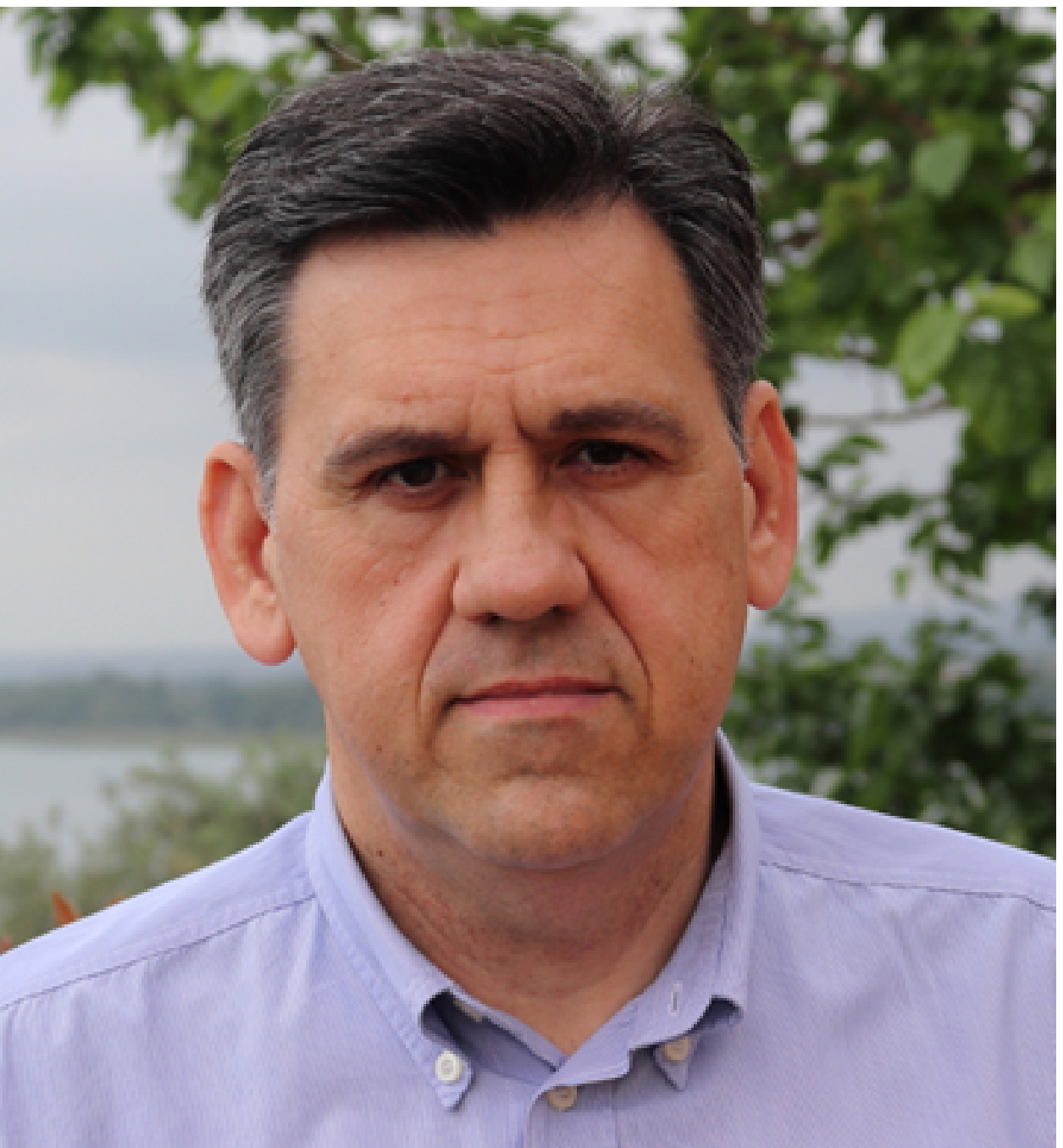}}]{Athanasios (Thanasis) G. Kanatas} (S'90-M'93-SM'02) is a Professor at the Department of Digital Systems, UNIPI, Greece. He received the Diploma in Electrical Engineering from the National Technical University of Athens (1991), the M.Sc. degree in Satellite Communication Engineering from the University of Surrey, UK (1992), and the Ph.D. degree in Mobile Satellite Communications from NTUA (1997). He has published more than 200 papers in international journals and conference proceedings. He is the author of 6 books in the field of wireless and satellite communications. He has been the technical manager of several European and National R\&D projects. His current research interests include the development of new waveforms and digital techniques for next generation wireless systems; wireless channel characterization and modeling; antenna design and array beamforming for next generation wireless systems; V2V communications and cybersecurity issues for V2V communications. He has been a Senior Member of IEEE since 2002. In 1999, he was elected Chairman of the Communications Society of the Greek Section of IEEE. From 2013 to 2017, he has served as Dean of ICT School of the University of Piraeus, Greece.
\end{IEEEbiography}

\begin{IEEEbiography}[{\includegraphics[width=1in,height
=1.25in,clip,keepaspectratio]{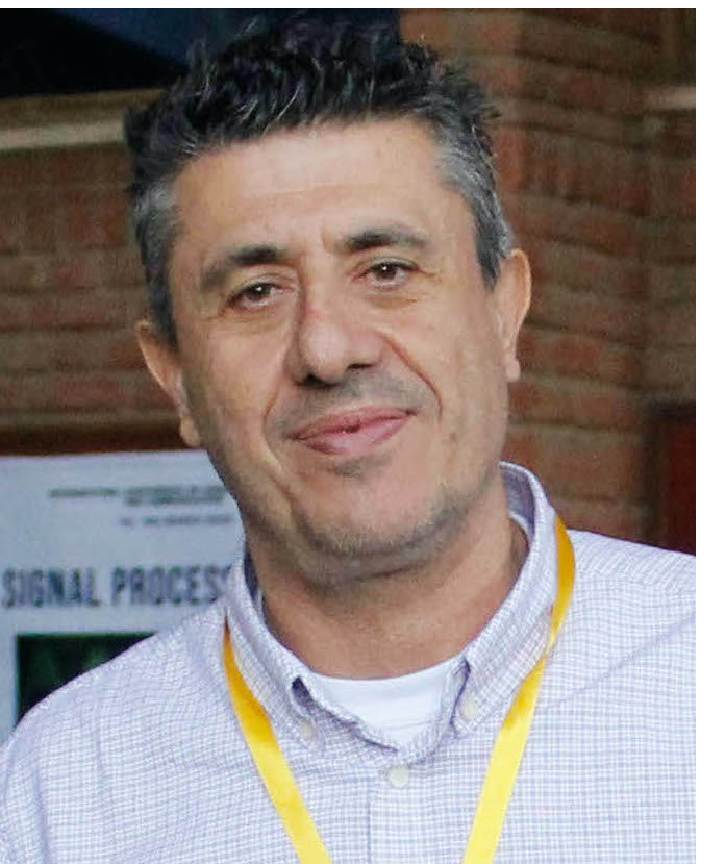}}]{George K. Karagiannidis} (M’96-SM’03-F’14) was born in Pithagorion, Samos Island, Greece. He received the University Diploma (5 years) and PhD degree, both in electrical and computer engineering from the University of Patras, in 1987 and 1999, respectively. From 2000 to 2004, he was a Senior Researcher at the Institute for Space Applications and Remote Sensing, National Observatory of Athens, Greece. In June 2004, he joined the faculty of Aristotle University of Thessaloniki, Greece where he is currently Professor in the Electrical \& Computer Engineering Dept. and Head of Wireless Communications Systems Group (WCSG).  He is also Honorary Professor at South West Jiaotong University, Chengdu, China.
His research interests are in the broad area of Digital Communications Systems and Signal processing, with emphasis on Wireless Communications, Optical Wireless Communications, Wireless Power Transfer and Applications and Communications \& Signal Processing for Biomedical Engineering. Dr. Karagiannidis has been involved as General Chair, Technical Program Chair and member of Technical Program Committees in several IEEE and non-IEEE conferences. In the past, he was Editor in several IEEE journals and from 2012 to 2015 he was the Editor-in Chief of IEEE Communications Letters. Currently, he serves as Associate Editor-in Chief of IEEE Open Journal of Communications Society. Dr. Karagiannidis is one of the highly-cited authors across all areas of Electrical Engineering, recognized from Clarivate Analytics as Web-of-Science Highly-Cited Researcher in the five consecutive years 2015-2019.
\end{IEEEbiography}

\end{document}